\documentclass[prl,showpacs,preprintnumbers,amsmath,amssymb]{revtex4}
\newcommand{\ket}[1]{\left | #1 \right\rangle}
\newcommand{\bra}[1]{\left \langle #1 \right |}

\usepackage[dvips]{graphics}

\begin{document}

\title{Interference in quantum field theory: detecting ghosts with phases}

\author{Chiara Marletto and Vlatko Vedral}
\affiliation{Clarendon Laboratory, University of Oxford, Parks Road, Oxford OX1 3PU, United Kingdom and\\Centre for Quantum Technologies, National University of Singapore, 3 Science Drive 2, Singapore 117543 and\\
Department of Physics, National University of Singapore, 2 Science Drive 3, Singapore 117542}

\date{\today}

\begin{abstract}
We discuss the implications of the principle of locality for interference in quantum field theory. As an example, we consider the interaction of two charges via a mediating quantum field and the resulting interference pattern, in the Lorenz gauge. Using the Heisenberg picture, we propose that detecting relative phases or entanglement between two charges in an interference experiment is equivalent to accessing empirically the gauge degrees of freedom associated with the so-called ghost (scalar) modes of the field in the Lorenz gauge.  These results imply that ghost modes are measurable and hence physically relevant, contrary to what is usually thought. They also raise interesting questions about the relation between the principle of locality and the principle of gauge-invariance. Our analysis applies also to linearised quantum gravity in the harmonic gauge, and hence has implications for the recently proposed entanglement-based witnesses of non-classicality in gravity. 
\end{abstract}


\maketitle                           

\section{Introduction}

The {principle of locality} is one of the pillars of the current best explanations of physical reality. It seems essential for any testable theory describing the universe as made of independent subsystems; and for defining concepts such as observables and interactions. 

`Locality' has a variety of meanings in the literature, but here `{\sl principle of locality}' shall mean, specifically, the principle that it must be possible to formulate any viable physical theory so that the state of any system made of two subsystems $A$ and $B$ is specified by a fixed function of an ordered pair of attributes $(a,b)$, where $a$ is an attribute of subsystem $A$ and $b$ is an attribute of subsystem $B$; and any transformation on $A$ can only change $a$ and not $b$. 

General relativity satisfies the principle of locality, as a consequence of Lorentz-invariance on which special relativity is based. Even non-relativistic quantum theory satisfies this principle -- as if it ``knew" about relativity in some roundabout way. This is because any state of a bipartite quantum system can be specified by an ordered pair of descriptors $(a,b)$, where $a$ and $b$ represent each a minimal set of generators for the algebra of local quantum observables. Hence the overall state of any bipartite quantum system is identified by the ordered pair $(a,b)$ and by a (constant) Heisenberg state $\psi$. The latter is a constant of motion specifying a ``fixed function" of $(a,b)$, as prescribed by the principle. Interestingly, this is true also when the two subsystems are entangled and display what is sometimes called (misleadingly) ``Bell-non-locality" -- see e.g. \cite{DEU}. 

Theories that satisfy the principle of locality in this form also satisfy the {\sl no-signalling} principle: this is the requirement that in a bipartite system the {\sl locally measurable properties} of one subsystem cannot be changed by transformations operating on the other subsystem only. The no-signalling principle is also satisfied by special and general relativity; and by quantum field theory (where it is sometimes also called ``microcausality" or ``causality"), as well as by non-relativistic quantum theory. Note that the speed of light plays no special role in determining whether a theory satisfies the principle: what is ruled out by the no-signalling principle is an instantaneous change of the locally measurable features of one system by acting on another one. This implies, for theories that admit a notion of velocity, that changes must propagate dynamically at some finite speed, which may or may not be the speed of light (see for instance the Lieb-Robinson bound on the propagation of dynamical perturbations in solid-state physics with finite-range interactions, \cite{LIE}). 

In field theory (both classical and quantum), to enforce the principle of no-signalling it is sufficient to require any allowed Lagrangian density to depend only on fields and no higher than the first derivatives of those fields, \cite{WEI}. This ensures that from one instant of time to the next, the propagation of disturbances in the field can only affect the observable features of nearest-neighbouring points (if Lorentz covariance is imposed, in addition, then this propagation happens at the speed of light). 

Assessing whether quantum field theory with sources satisfies the principle of locality, however, is less straightforward. In the case of quantum electrodynamics there is no known way to describe point-like interactions between local observables of subsystems by referring to the electric and magnetic fields only, \cite{COH, DEW}. There exist accounts in terms of such fields only, but they involve global quantities, as for instance discussed in \cite{DEW}. In classical physics instead it is always possible to provide a complete, local description of all phenomena using the electric and magnetic fields only. This is because the classical formulation can be fully expressed by locally appealing to forces on charges, exerted directly by the electric and the magnetic fields; but in the quantum case this does not seem to be possible. Fortunately, in the quantum case it is still possible to have a local description in terms of point-like interactions by resorting to the so-called ``potentials" as local quantum descriptors of the field. (A classical Lagrangian and Hamiltonian formulation of the interaction between charges and fields also goes via the potentials, but this is not a necessity as the local force-based account always exists, so potentials are not necessary.)
The problem with potentials, however, is that they are gauge-dependent quantities, and the principle of locality (as defined above) is not satisfied in all gauges. In fact, the gauges that satisfy locality form a class, and the Lorenz gauge is its best-known representative (see e.g. \cite{MUL} for a review). 

Does this mean that the Lorenz gauge is more physical, more ``real", than others, given that it obeys the locality principle? Here is where the issue becomes even more interesting. As we shall explain, the Lorenz gauge resorts to four modes in order to achieve its locality: two transverse, one longitudinal, and one scalar. The latter two modes, which exist as a consequence of enforcing Lorentz-covariance (as they form, with the transverse modes, a $4$-vector), are associated with problematic mathematical features (such as negative norms, \cite{COH}).  It is well-known that these problems can be corrected for by imposing particular supplementary conditions on the allowed states of the electromagnetic field, for example following the Gupta-Bleuler prescription, \cite{COH}. The supplementary conditions, as we shall discuss later, make those modes not directly observable, and hence they are usually called ``ghosts". 

In this paper however we refute of the idea that ``ghost modes" are unobservable. To this end, we shall analyse a local model for the generation of a relative phase between two charges via a quasi-static Coulomb interaction, in the Lorenz gauge.  This regime corresponds to two slowly-moving charges, each with the ability to be superposed across different locations, interacting via the Coulomb potential. Our results are readily applicable to linearised quantum gravity in the quasi-static Newtonian regime, in the harmonic gauge, which is equivalent to the electromagnetic case. Interestingly, this model is the basis for the recently proposed experiment to test non-classicality of the gravitational field via the generation of gravitational entanglement between two masses, \cite{MAVE, SOUG, MAVE2, MV-PRD}: the relative phase is responsible for the entanglement.  

We shall show that, both in the electromagnetic and the gravitational cases, the relative phase formation can be described by resorting to the scalar ghost modes of the Lorenz gauge, in a way that fully complies with the principle of locality (as defined earlier). We shall then point out the interesting fact that measuring relative phases or the entanglement between the charges is equivalent to providing a tomographic reconstruction of the q-numbers representing the scalar ghost modes of the electromagnetic field. This is evident using the Heisenberg picture of quantum field theory. 
Hence, contrary to what is usually thought, ghost modes (though not directly excitable) can be indirectly detected by measuring the quantum phases that they mediate. 

This result could be interpreted in different ways. First, it ignites an interesting theoretical discussion about the validity of the quantum theory of fields as it is currently formulated; if we insist on the latter to be viable, and we take the results in this work seriously, we must conclude that there are entities (such as the ghost modes) that can be indirectly observed but are not necessarily represented by Hermitean operators or with positive-norm eigenstates. This may suggest that the current formulations of quantum field theory with potentials are inadequate as they are internally inconsistent.  

In addition, it provides a new argument in support of the conjecture that the Lorenz gauge provides the only complete description of reality, which fully complies with the locality principle. 

Another interesting implication is that the principle of locality (as formulated above) and the principle of gauge-invariance are not in contradiction with each other; however, at least in the current formulations of quantum field theory with interactions, the locality principle is satisfied only in a particular class of gauges.  

There are a few other options that we outline in the final discussion, but all these options seem radical; they all hint at the fact that the information-theoretic structure of quantum field theory requires a deeper understanding than we currently have.

\section{A local description of the quantum interaction between two charges}

We shall now describe the standard quantum field theory treatment of the interaction between two identical charges $q$ in the Lorenz gauge, following the treatment in \cite{COH}, which is based on the Gupta-Bleuler construction. We shall analyse the electromagnetic case, but the results can be straightforwardly extended to the gravitational case in linear quantum gravity models, \cite{GUP1, GUP2}. 

We shall consider a slowly-varying charge distribution (e.g. with one charge undergoing interference in the potential generated by the other charge) and so we shall adopt the adiabatic approximation, focussing only on the Hamiltonian describing the interaction of the two charges placed at two fixed locations. By linearity, we can then infer the relative phase of one superposed charge with respect to the other, and finally how (when both 
charges are superposed) they become entangled. We shall also assume the metric tensor corresponding to Minkowski flat spacetime, $g_{\mu\nu}$ : $g_{00}=+1,  g_{11}=g_{22}=g_{33}=-1$.

Even if we are restricting attention to a static Hamiltonian, note that the situation we are considering is non-static: to create and detect the relative phase between two charges, it is necessary for at least one of the charges to dynamically change from the configuration where its position is sharp, to the configuration where it is not, and back again. We shall not describe explicitly this dynamical process, because (for the purpose of arguing for the measurability of the ghost modes) it is sufficient to restrict attention to the Hamiltonian describing the interaction between the static charges in the adiabatic approximation.

We shall use ${\vec r_i}$ to denote a particular 3-vector representing the position of a charge.  We shall also denote by $b_i$, $b_i^{\dagger}$ the bosonic annihilation and creation operators relative to the charge located at ${\vec r_i}$ {(in the fully relativistic regime these should be fermionic operators representing a Dirac field, \cite{COH}, but in this regime they can be approximated by spinless bosonic particles).} The charge density operator for a two-charge distribution is given (as a function of the position 3-vector $\vec r$) as

\begin{equation}
\hat \rho ({\vec r_{ij}}) = q \left (b_i^{\dagger}b_i\delta({ \vec r}-{ \vec r_i})+b_j^{\dagger}b_j\delta ({ \vec r}-{ \vec r_j}) \right )\;.
\end{equation}

Its Fourier transform in momentum space (as a function of the 3-vector $\vec k$ representing the 3-momentum) is written as:

\begin{equation}
\hat\rho_{ij} (\vec {k}) = \frac{1}{{(2\pi)}^{\frac{3}{2}}}q (b_i^{\dagger}b_i\exp{(-i{ \vec k\cdot \vec r_i})}+b_j^{\dagger}b_j\exp{(-i{ \vec k \cdot\vec r_j})})\;.
\end{equation}

The total Hamiltonian in the Lorenz gauge is $$H_{L}=H_F+H_I\;,$$ where $H_F$ is the free Hamiltonian of the field and $H_I$ is the interaction Hamiltonian. 

The free Hamiltonian $H_F$ is written as:

\begin{equation}
H_F= \int {\rm d}^3k\hbar \omega(k) \left(\sum_{\nu=1}^{3}a_{\nu}(\vec k)^{T}a_{\nu}(\vec k) -  a_{0}(\vec k)^{T}a_{0}(\vec k)\right)\;.
\end{equation}

Here, $a_{\mu}(\vec k)$ is the annihilation operator for the k-th mode of $\mu$ component of the electromagnetic field, where $\mu=0, 1, 2, 3$, while $a_{\mu}(\vec k)^{T}$ is its adjoint; and we have defined $\omega(k)=ck$, ($k= |\vec k|$). For the scalar modes, i.e. when $\mu=0$, the adjoint operation $^T$ does {\sl not} correspond to the standard adjoint and shall be defined later; while for all the other modes, it is the standard $\dagger$ adjoint operation.  
As mentioned, in order to achieve Lorentz-covariance four modes must be present in the Lorenz gauge: three spatial  -- i.e., two transverse ($\mu=1,2$), one longitudinal ($\mu=3$) -- and one temporal or scalar ($\mu=0$). Note that this Hamiltonian is different from the standard free field Hamiltonian for the electromagnetic field (which only includes the transverse modes): it is obtained from a modified Lagrangian designed to include the vector potential as a four-vector, with each component having a non-zero conjugate momentum.

Assuming the metric tensor to be that of a flat Minkowski spacetime implies that the creation and annihilation operators for the scalar modes must obey the modified commutation relation: $$[a_{0}(\vec k), a_{0}(\vec k)^{T}]=-1\;.$$ 

As a consequence, the scalar odd number states $\left (a_{0}(\vec k)^{T}\right)^{2n+1}\ket{0}$ have a negative norm if the norm is defined as usual and $^T$ is assumed to denote the usual adjoint operation. This problematic issue is frequently glossed over because, in the Lorenz gauge, a particular {\sl supplementary condition} on the allowed states must also be satisfied, in order to have consistency with Maxwell's equations, \cite{SHA, COH}. This condition in particular ensures that Gauss' law is satisfied at all times during the unitary dynamical evolution by simply annihilating the states that do not satisfy it. For the free field, the condition is the quantum equivalent of the classical constraint defining the Lorenz gauge, $\sum_{\mu}\partial_{\mu}A_{\mu}=0$, $A_{\mu}$ being the classical 4-vector potential. This classical condition cannot be satisfied in quantum theory (as the potential components are q-numbers), but it can be expressed by requiring that the allowed states satisfy the eigenvalue equation:

$$
(a_{3}(\vec k)-a_{0}(\vec k))\ket{\psi}=0, \;\forall \; \vec k \;.
$$

Supplementary conditions of this kind, as we shall see, make the states sharp with any number of excitations of the scalar and longitudinal modes unobservable. Hence, as we said, they are traditionally called ghost modes (and their excitations are called ``virtual" particles). 

In combination with this supplementary condition, one can adopt some measures to render the theory mathematically viable despite the presence of a negative norm. One possible approach is the Gupta-Bleuler construction. It resorts to a different metric, \cite{COH}, which we shall call $M$-metric. The creation operator of the scalar photons are defined by a modified ``M-adjoint'' operation, with the property that: $a_{0}(\vec k)^{T}\doteq  M a_{0}(\vec k)^{\dagger}M$, where $[a_{0}(\vec k), a_{0}(\vec k)^{\dagger}]=1$ are standard creation and annihilation operators for a harmonic oscillator and $M$ is a unitary, self-adjoint operator, with the property that $M a_{0}(\vec k)^{\dagger}M= - a_{0}(\vec k)^{\dagger}$. Informally, this unitary inverts the sign of odd-numbered states and leaves the even-numbered unchanged, thus addressing the negative norm issue. It can also be represented as the unitary parity phase operator $(-1)^{a_{0}(\vec k)^{\dagger}a_{0}(\vec k)}$. The new inner product of two general states in this ``M-metric" is then defined as: $$\langle\langle\psi|\phi\rangle\rangle\doteq \bra{\psi}M\ket{\phi}\;.$$ With this new inner product one can check that the $M$-norm of odd-numbered states is positive, as expected. We shall adopt the $M$-metric throughout the calculation of the phase for the scalar photon sector. 

The interaction Hamiltonian $H_I$ is written as:
\begin{equation}
H_I= \sum_{\mu}\int {\rm d}^3r\hat j_{\mu}\hat A^{\mu}\;,
\end{equation}
where in our case $\hat  j_{\mu}= (\hat \rho ({\vec r_{ij}}), 0, 0, 0)$ is the source four-vector and $\hat A^{\mu}$ is the potential four-vector operator, with $A^0$ being the scalar potential.  By substituting the expression for the quantised scalar potential in momentum space and for the charge distribution $\hat \rho(\vec r_{ij})$, we find:

\begin{equation}
H_I= \int {\rm d}^3 k qc\sqrt{\frac{\hbar}{2\epsilon_0\omega(k)(2\pi)^3}}[a_{0}(\vec k) \left (b_i^{\dagger}b_i\exp{(-i\vec k \cdot\vec r_i)}+b_j^{\dagger}b_j\exp{(-i\vec k\cdot\vec r_j)}\right )+ {\rm adj.}]\;, \label{HAM}
\end{equation}
where `${\rm adj.}$' denotes the $^T$ operation on the scalar photons sector and the usual adjoint for all other operators. 
In the presence of interactions, the supplementary condition is modified to include the charge distribution operator, as:

\begin{equation}
(a_{3}(\vec k)-a_{0}(\vec k)+\hat \eta_{ij}(\vec k))\ket{\psi}=0, \forall \vec k \label{supp}
\end{equation}
where we have defined the operator $\hat \eta_{ij}(\vec k)= c\sqrt{\frac{\hbar}{2\epsilon_0\omega(k)(2\pi)^3}}\hat \rho_{ij}(k)$.

It is often said that the Coulomb interaction is instantaneous, or even non-local. We note, however, that this is incorrect (as already pointed out by many, e.g. \cite{COH}). In particular, in the Lorenz gauge, the Coulomb interaction Hamiltonian $H_I$ satisfies the locality principle. For if one considers the action of each of its terms on each of the charges and the field, the subsystems whose descriptors do not appear in that term are not affected by that term directly. In particular there is no direct, instantaneous interaction between the two charges, only a mediated interaction via the field. This is an important feature of the Lorenz gauge which makes it satisfy the locality principle. 

The process governed by the interaction Hamiltonian $H_I$ is sometimes described as an ``exchange of virtual photons" (i.e. ghost photons) in which one charge ``emits" a virtual photon, which then propagates (at the speed of light, and in a perfectly causal way) to the other charge where it is absorbed. It is impossible, however, to think of the virtual photons as independent of the charges; unlike the photons of the transverse modes in the free electromagnetic field, these virtual photons cannot exist without charges. Hence the expression ``virtual photons'' is misleading in the same way that it is misleading to say that atoms bond into molecules by ``exchanging" electrons. Electrons do not hop between atoms, they simply just go into the new collective eigenstates of the bonding atoms, and the lower energy of these eigenstates (compared to the energies of the individual atoms) is the quantum ``force" behind chemical bonds. The same is true of the Coulomb and Newton ``forces" in quantum field theory. In an almost perfect analogy with chemistry, a positive and a negative charge attract each other electromagnetically by sharing scalar photons whose new eigenstates - when two charges are present - have lower energy than the state they would have when pertaining to each charge individually. The analogy seemingly breaks down only because the shared electrons in molecules can be detected directly, while the shared (scalar) photons in quantum field theory are allegedly undetectable. However, it is exactly this point that we question next.

\section{The Schr\"odinger picture}

Here we discuss how the Hamiltonian leads to a detectable relative phase when one of the two charges is superposed across different spatial modes. We can perform the calculation in a number of ways, but we shall follow the construction outlined in \cite{SHA, COH}. 

Let us focus on the scalar part of the Hamiltonian only, as it is the only one relevant for this problem (we shall therefore drop the index $0$ in the photon creation and annihilation operator). We can write the scalar Hamiltonian as $H=\int {\rm d}^3 k\hbar \omega(k) (H_{fk} +H_{Ik})$, where

\begin{equation}
H_{fk}=- a(\vec k)^{T}a(\vec k)\;,
\end{equation}

\begin{equation}
H_{Ik}= [a(\vec k) \hat\eta_{ij}(\vec k)^{\dagger}+\hat\eta_{ij}(\vec k) a(\vec k)^{T}] \;.
\end{equation}

We can also introduce the notation $\hat\eta_{ij}(\vec k)= \frac{g(k)}{\hbar\omega(k)}(b_i^{\dagger}b_i\exp{(-i\vec k \cdot \vec r_i)}+b_j^{\dagger}b_j\exp{(-i\vec k \cdot\vec r_j)})$, where $g(k)=qc\sqrt{\frac{\hbar}{2\epsilon_0\omega(k)(2\pi)^3}}$. Note that the $-$ sign in $H_{fk}$ is the signature of scalar photons: it is a consequence of requiring that the Lagrangian generating this Hamiltonian be Lorentz-covariant and the metric be Minkowski, \cite{HAL, Deser}. 
The supplementary condition for the Lorenz gauge is expressed by requiring that the allowed states belong to the subspace defined by \eqref{supp}. Since the longitudinal photons are in the vacuum state for this particular problem, we can omit them from the description. Hence the condition \eqref{supp} implies that the allowed states for the scalar photons and the charges is the span of these states:

\begin{equation}
\ket{\psi}=\ket{c_{ij}}\ket{\lambda_{\vec r_i,\vec r_j}}_0\;.\label{HS}
\end{equation}

Here, the charge sector is described by the state $\ket{c_{ij}}=b_i^{\dagger}b_j^{\dagger}\ket{0}_c$ ($\ket{0}_c$ being the vacuum state of the charge field), which represents the state where one charge is located at position $\vec r_i$ and the other at position $\vec r_j$; while $\ket{\lambda_{\vec r_i,\vec r_j}}_0$ is an eigenstate of the scalar photon operator $a_{k}$ with eigenvalue $\lambda_{\vec r_i,\vec r_j}= \frac{g(k)}{\hbar\omega(k)}(\exp{(-i\vec k\cdot\vec r_i)}+\exp{(-i\vec k\cdot\vec r_j)})$ (which is the eigenvalue of $\hat \eta_{ij}(\vec k)$ in the state $\ket{c_{ij}}$). 

It is possible to diagonalise the above Hamiltonians using the linear transformation:
$$D(\vec k)= \exp( \hat\eta_{ij}(\vec k)a(\vec k)^{T}- a(\vec k)\hat\eta_{ij}(\vec k)^{\dagger} )\;,$$
which can be proven to be a displacement operator on the scalar photon field and unitary with respect to the $M$-norm (i.e., $D(\vec k)^{T}D(\vec k)=id=D(\vec k)D(\vec k)^{T}$), \cite{COH}. Note that, due to the different commutation relations of the scalar photons, we have:
$$D(\vec k)^{T} a(\vec k) D(\vec k)= a(\vec k)+\hat\eta_{ij}(\vec k) $$
$$D(\vec k)^{T} a(\vec k)^{T} D(\vec k)= a(\vec k)^{T}+\hat\eta_{ij}(\vec k)^{\dagger} $$

$$D(\vec k)^{T} a(\vec k)^{T}a(\vec k) D(\vec k)= a(\vec k)^{T}a(\vec k)+ \hat\eta_{ij}(\vec k)^{\dagger} a(\vec k)+\hat \eta_{ij} a(\vec k)^{T}+\hat\eta_{ij}(\vec k)^{\dagger} \hat\eta_{ij}(\vec k)\;.$$

The diagonalised Hamiltonian $H(\vec k)^{d}$ reads:

$$H(\vec k)^{d}=D(\vec k)^{T}H(\vec k)D(\vec k)= \hbar \omega(k) (-a(\vec k)^{T}a(\vec k)+\hat\eta_{ij}(\vec k)^{\dagger}\hat\eta_{ij}(\vec k))\;.$$

\noindent By integrating over $\vec k$, we see that the phase generated by $U=\exp(-iHt)$ acting on the initial state set by the supplementary condition is:

$$\langle\bra{\psi_H}U\ket{\psi_H}\rangle= \exp(-i\phi_c)\;,$$

\noindent where $\phi_c(\vec r_i, \vec r_j)= \int {\rm d}^3k\hbar \omega(k)\lambda_{\vec r_i,\vec r_j}^{*}\lambda_{\vec r_i,\vec r_j}$, and we have used the fact that $\exp(-i H(\vec k)t)=D(\vec k)\exp(-i H(\vec k)^{d}t)D(\vec k)^{T}$ and that, by definition of displacement operator, $D(\vec k)^{T}\ket{c_{ij}}\ket{\lambda_{\vec r_i,\vec r_j}}_0=\ket{c_{ij}}\ket{0}_0$, where $\ket{0}_0$ is the vacuum of the scalar field.

By noting that $\lambda_{\vec r_i,\vec r_j}^{*}\lambda_{\vec r_i,\vec r_j} =2 \left(\frac{g(k)}{\hbar\omega(k)}\right)^2 (1 + \cos(k(\vec r_j - \vec r_i))$, by integration in spherical coordinates (see e.g. \cite{MUE}) we find the familiar Coulomb phase:

$$\phi_c(\vec r_i, \vec r_j)= \frac{q^2}{4\pi \epsilon_0 \hbar |\vec r_i-\vec r_j|}t\;,$$

(up to a constant term not dependent on the position).

This phase is unobservable when both charges have a sharp position since it is then only a global phase. However, when one of the two charges is superposed across two locations, we obtain a relative phase between the two branches of the quantum state of the charge, that can be measured by performing interference on that charge. This reveals the presence of the other charge, whose field generates the (relative) phase. 

\section{The Heisenberg picture}

In the Schr\"odinger picture (and in treatments such as the S-matrix approach, \cite{WEI}), the phase is calculated as a global property of the quantum state describing the charges and the electromagnetic fields - hence it is not possible to see explicitly how the degrees of freedom of the scalar photons are responsible for the phase formation on the charges' subsystem. In the Heisenberg picture, we can instead calculate the change of the q-numbers describing each subsystem of the charge and the field: this shall give us an interesting local perspective on the phase, showing us how detecting the relative phase on a charge due to the presence of another charge amounts to indirectly measuring the degrees of freedom of the scalar photon field. 

First, consider the dynamical evolution of the q-numbers describing the scalar electromagnetic field. For each $\vec k$, the generators of the local algebra for each mode evolve as follows:

$$a(\vec k)(t)= U^{T}a(\vec k)U= \exp(-i\hbar\omega(k) t) (a(\vec k)-\hat\eta_{ij}(\vec k))$$

$$a(\vec k)^{T}(t)a(\vec k)(t)= (a(\vec k)^{T}a(\vec k)-\hat\eta_{ij}(\vec k)^{\dagger}a(\vec k) -\hat \eta_{ij}(\vec k) a(\vec k)^{T} +\hat\eta_{ij}(\vec k)^{\dagger}\hat\eta_{ij}(\vec k) )\;,$$

where $U^{T}$ is the standard adjoint of the operator $U$ on the charge sector, while on the photon sector it is defined as the $M$-adjoint operation introduced earlier. Hence all expected values of the quadrature $a(\vec k)(t)+a(\vec k)^{T}(t)$ and of the number operator $a(\vec k)^{T}a(\vec k)(t)$ are zero in the Heisenberg states $\ket{\psi}$ that satisfy the supplementary condition \eqref{HS}. This is consistent with the interpretation that these scalar modes are undetectable. Yet, as we can see, they are necessary to account for the local generation of the phase. We shall now challenge the position that they are undetectable, because, as we shall argue, their degrees of freedom can in fact be indirectly accessed experimentally by measuring the properties of the charge, in particular the relative phase they mediate. 

To see how, we can consider the change induced in the q-numbered observables of one charge due to the mediated interaction with the other charge. We first consider how each of the generators of the algebra of observables of the charge is modified by the Hamiltonian $H$:

\begin{equation}
b_i(t)= U^{T}b_iU= \exp(-i\hat \alpha_i t) b_i \label{charge}
\end{equation}

where we introduced the self-adjoint operator $\hat \alpha_i= \int {\rm d}^3k \frac{g(k)}{\hbar}(a^{T}\exp{(-i\vec k\cdot\vec r_i)}+h.c.)$. Hence while the number operator for each charge stays invariant, there are more general, joint observables of both charges get modified by the interaction Hamiltonian. These joint observables are relevant for the relative phase between one charge and the other, and for any entanglement between the two. 
We shall now demonstrate how measuring these observables corresponds to measuring indirectly the scalar degrees of freedom, using a single charge superposed across two locations, in the presence of a second charge (at another location). 

\subsection{Detecting ghosts indirectly using the relative phase}

It is easy to see that the observables of the charge end up depending on the degrees of freedom of the ghost field, via the action of the Hamiltonian $H$, even if these degrees of freedom are declared to be unobservable by standard quantum field theory interpretations. Hence, under a more general definition of observable -- i.e., any degree of freedom that is copiable, \cite{MV-PRD}-- we can claim that these ghost degrees of freedom count as such, in the sense that they can be `copied' into the charge's degrees of freedoms which in turn can be measured. 

Let us consider the situation where a charge $A$ is superposed across two spatial modes, $\vec r_i$ and $\vec r_m$, and another charge $B$ is at a fixed position $\vec r_j$. 

The charge observable relevant for the relative phase is $C_{A}=b_i^{\dagger}b_m + b_m^{\dagger}b_i$: if the charge A is superposed across two different locations, $\vec r_i$ and $\vec r_m$, this observable is sharp with value 1. 

The interaction with the charge $B$ at $\vec r_j$ causes the following modifications to the observable $C_A$, according to equation \eqref{charge}:

\begin{equation}
C_A(t)= U^{T}C_AU= \exp(-i(\hat \alpha_i-\hat \alpha_m)t) b_i^{\dagger} b_m+{\rm adj.}
\end{equation}

When there is one charge at $\vec r_i$ and the other at $\vec r_j$, the supplementary condition \eqref{HS} is $\ket{\psi}=\ket{c_{ij}}\ket{\lambda_{\vec r_i,\vec r_j}}_0$;  while the case where the charge $A$ is at location $\vec r_m$, and the other charge at $\vec r_j$, corresponds to the state $\ket{\psi}=\ket{c_{mj}}\ket{\lambda_{\vec r_m,\vec r_j}}_0$. Considering a Heisenberg state where the charge is in a superposition of those two states, e.g. $\ket{\psi_H}=\frac{1}{\sqrt{2}} (\ket{c_{ij}}\ket{\lambda_{\vec r_i,\vec r_j}}_0+\ket{c_{mj}}\ket{\lambda_{\vec r_m,\vec r_j}}_0)$ one can compute the expected value of $C_A(t)$, which results to be a function of the relative Coulomb phase between the two branches $\vec r_i$ and $\vec r_m$ of the quantum state of the charge $A$, caused by the Coulomb interaction with the other charge at location $\vec r_j$. So we can measure the Coulomb relative phase by measuring the observable $C_A(t)$ of charge A, for example using local tomography on each spatial mode  $\vec r_i$ and $\vec r_m$, without closing the interference loop (see for instance the protocol discussed in \cite{MV-AB}).  

It is important at this point to notice that by measuring $C_A(t)$ we also indirectly measures the q-numbers $\hat \alpha_i$s, which include the degrees of freedom of the ghost field. Hence, we can consider this process a measurement of the ghost degrees of freedom, albeit indirect. We say `indirect' because the scalar photons are not the eigenstates of any Hermitian operators pertaining to those modes. The possibility of performing this measurement seems in contradiction with the traditional interpretation of quantum field theory, that ghosts are unobservable. 

This raises an interesting question, namely: could one prepare and detect, in the above indirect sense, other states of the scalar modes in addition to the coherent states $\ket{\lambda_{\vec r_i,\vec r_j}}_0$? If not, then the intriguing conclusion would be that the only indirectly detectable scalar states are coherent states, which are the most classical of all states of light (strictly speaking, only in the limit of the infinite amplitude are the coherent states classical, but we do not wish to go into this discussion here). 

\section{Discussion}

We have just shown that measuring the relative Coulomb phase without closing the interference loop  is equivalent to indirectly measuring variables pertaining to the ghost scalar modes. This is because in order to provide a local model of the generation of the Coulomb relative phase we need to invoke the scalar ghost modes in the Lorenz gauge.

The same conclusions can be applied to linear quantum gravity describing a quasi-static Newtonian interaction between two masses $m$, in the adiabatic approximation. One can express the relevant Hamiltonian with an analogous construction resorting to scalar modes in the harmonic gauge, following e.g. \cite{GUP2}. In order to adapt the results of our calculation, one needs to modify the relevant constants (i.e., by replacing $q$ with the mass $m$ of the two quantum systems and taking $g(k)=mc\sqrt{\frac{G}{\hbar\omega(k) (2\pi)^{2}}}$, $G$ being the gravitational coupling constant); and take into account that gravitational potential is represented by a tensor field, whose only relevant components in the quasi-Newtonian regime are the scalar ones. Hence, experiments where a single mass undergoes interference in a gravitational field, such as \cite{COW} and \cite{KAS}, could in principle be modified to include an indirect measurement of the scalar degrees of freedom of the field, if we resort to this model of linearised quantum gravity to describe them and we perform a tomographic reconstruction of the gravitational phase without closing the loop. Likewise for the experiments where gravitational entanglement is generated locally between two quantum superposed masses, \cite{SOUG, MAVE}. However, note that for the single-mass experiments to be used as evidence of the non-classicality of the gravitational field additional assumptions are needed compared to the entanglement-based witnesses, \cite{DIMA}. This is because for a semiclassical model (where the field is classical) is sufficient to provide a local account of the interference with a single charge; while it is not sufficient when entanglement is generated. 

Our result has interesting implications. One is that if one insists on having a local account of the generation and detection of the Coulomb phase, acquired by interaction between two quasi-static charges, then the Lorenz gauge (and gauges in the same equivalence class) emerges as the only one viable. Furthermore, by analysing the phase generation in the Heisenberg picture, we see that ghost modes have a certain degree of reality, in the sense that they can be indirectly observed, as we explained, by measuring the quantum phase on the superposed charge. Note that it is still true, as claimed in several accounts of quantum field theory, that the ghost modes cannot be excited: there is no creation or annihilation of a single photon during this interaction. However the degrees of freedom of the ghost photon field are engaged when explaining the phase in a local way, and they can be measured indirectly as we discussed. 

This fact makes us question the validity of the quantisation in the Lorenz gauge: on the one hand, the straightforward quantisation procedure leads to scalar operators having eigenstates with negative norms (or, equivalently, to non-Hermitean operators, or indefinite metric constructions, \cite{COH, SHA}; or non-unitary dynamical evolutions, \cite{HAL}). On the other, using the very same theory, we end up concluding that those degrees of freedom are observable after all. This seems to violate the central tenet that observables can be described as Hermitean operators (directly as such, or as a limiting point of sequences of well-defined Hermitean operators, like in the case of general quantum phases). After all, the most general quantum measurement on a given system entails coupling that system to another one and then measuring both in some, normally entangled, basis. Here, however, because of the impossibility of observing scalar number states, this generalised measurement account for the phase fails.  This hints at a contradiction in the present formulation of quantum field theory. 

A possible way to address this problem is to postulate that photons have a non-zero mass. This possibility has been considered from time to time, \cite{REV}: a non-zero photon mass would make the ghost photons real and it would eliminate the gauge redundancy. This could in principle have detectable consequences; however, if this mass is below a certain limit, it could still be experimentally inaccessible (due to the weakness of its coupling to matter). 

If instead the photon mass is identically zero, does our result mean that gauge-invariance is violated? No. Gauge-invariance (as intended in quantum field theory) requires global observable properties of the system of field and charges {\sl together} to be unchanged when a gauge transformation is applied, \cite{BRABRO}. This property is still true in the case we analysed. The phase $\phi_c$ is gauge-invariant in this sense: any gauge transformation, being an invertible transformation acting jointly on the observables and on the Heisenberg/initial state, leaves that phase invariant. (For the interested reader, in \cite{COH, FRA, HAL} it is possible to find the expression for the invertible transformation that allows one to map the Lorenz gauge description onto the Coulomb gauge description. This transformation involves both the charge and the photon field degrees of freedom.)

The dynamical account of how the phase is acquired, on the other hand, is not gauge-invariant. If we want to adhere to a local account, we need to select the Lorenz gauge (or equivalent ones) considering the current understanding of local interactions in quantum field theory. It is curious to note that this is very much in analogy with the different pictures of quantum theory. The Heisenberg, Schr\"odinger, and interaction picture are all consistent as far as the empirical content of their predictions. However, they give different dynamical accounts of how the observable effects are achieved - and only in the Heisenberg picture (like in the case of the Lorenz gauge) the description is explicitly local. In both cases, the existence of the explicitly local description (in the Lorenz gauge, and in the Heisenberg picture in the two respective cases) is what makes the theory in question (quantum field theory and quantum theory in the two cases) comply with the locality principle (as we defined it in the opening paragraph of the paper). 

It is worth also stressing that when in an interference experiment, the charge involved has to be in motion at least during the time when the superposition of its different spatial locations is prepared. No matter how slow this process is, it is still dynamical and therefore the longitudinal and the transverse modes of the field become excited. The longitudinal modes are ghosts too and hence longitudinal photons are undetectable, \cite{COH} -- a requirement that is implemented by the previously-mentioned supplementary condition.  Some scalar photons do become real when the charge is accelerated, but they only manifest themselves in the transverse modes. The transverse photons can then be detected, but their amplitude is so small, due to the adiabaticity of the process, that they clearly have no relevant effect here. 

What does this result imply for the witnesses of non-classicality in gravity, and related experiments? Such witnesses require one to double the interference setup described in the previous section, with two charges superposed each across two different locations interacting with each other. In that case too, by our reasoning, confirming entanglement is equivalent to indirectly measuring the degrees of freedom of the gravitational (scalar) field as described in linear quantum gravity models. Note that one can place this interpretation on observing entanglement once a particular model of linear quantum gravity is selected - observing entanglement by itself does not necessarily confirm this linear quantum gravity model as opposed to others, it simply allows us to refute all classical theories in a given class (see \cite{MV-PRD} for a discussion of this point). 

Our analysis also shows the importance of assuming locality in the tests of non-classicality of gravity. Without the principle of locality, there is no way to argue that the mediator of entanglement is quantum, as there is no such thing as an independent mass or subsystem. The static interactions like Coulomb's and Newton's are indeed sometime presented as non-local; this has led to claims that detecting gravitational entanglement is not a conclusive proof of the non-classicality of gravity, \cite{BEI}. In such accounts, the charge (or mass) is dressed with the scalar mode first which effectively makes the charge an extended object. When two charges are present, the cloud of photons now has the possibility to be in a superposition state between the two charges which has a higher (or lower) energy than the charges on their own. It is this change in energy that manifests itself as the Coulomb repulsion (or attraction when the charges are opposite). This kind of account embodies the philosophy that the ghost photons do not have an independent existence and should be seen as part of the charges. However, in this account locality is lost (though the account is still perfectly causal - it satisfies the no-signalling principle). If on the other hand one assumes the locality principle and detects gravitational entanglement generated by local means, then it is possible to conclude that the mediator of the entanglement is non-classical; and in the model we just discussed it consists of the scalar modes of the gravitational field. 

Another possible way of interpreting our results is to reject the principle of locality. This however seems to be at odds with various aspects of our understanding of physical reality. Universally accepted non-local theories do not exist at present, but many have been constructed and discussed (e.g. Bohmian mechanics or Feynman-Wheeler absorber theory). It could be interesting to analyse these processes in those theories - with the caveat that in order to set up an interference experiment or an entanglement verification, one already usually assumes locality. It is our conjecture that locality shall remain a key property for all testable theories.

We would like to close with a different way of expressing our dissatisfaction with the present state of quantum field theory. In a paper in honour of Bohm, Feynman \cite{Feynman} considered the notion of negative probabilities in quantum physics. The point he made was that, if we allowed negative probabilities, we would not need entanglement to violate Bell's inequalities; local hidden variables would suffice, so long as their averaging is done with both positive and negative probabilities. 

This solution would, of course, not be acceptable to most physicists, because we think of probabilities as limiting points of frequencies of observed outcomes on repeated experiments -- and such frequencies, or their limits, cannot be negative.  Likewise, in quantum field theory negative norms in the scalar modes could be viewed as negative probabilities (something that Feynman uses to motivate his exploration in the aforementioned paper) and therefore ought to be regarded as equally problematic. But this is usually not considered as a serious problem, because the photons in these modes are supposed to be virtual and not directly detectable (Feynman too in his analysis proposes negative probabilities to be unobservable). In this paper we have challenged the idea that scalar photons are unobservable degrees of freedom, and argued that they can be detected through interference of charges (unlike negative probabilities) -- for example as in the gravitational witnesses of non-classicality, where the entanglement detected between the masses is (almost completely) due to the scalar modes. Coupled with the principle of locality, these considerations should make the ``reality" of the scalar modes beyond doubt; and should urge us to reconsider the foundations of quantum field theory. 

\textit{Acknowledgments}: The Authors thank Abhay Ashtekar and David Deutsch for suggestions and comments. CM thanks the Eutopia foundation. VV thanks the National Research Foundation and the Ministry of Education, Singapore, for funding this research as administered by Centre for Quantum Technologies, National University of Singapore. This publication was made possible through the support of the ID 61466 grant from the John Templeton Foundation, as part of the The Quantum Information Structure of Spacetime (QISS) Project (qiss.fr). The opinions expressed in this publication are those of the Authors and do not necessarily reflect the views of the John Templeton Foundation.

\end{document}